\documentclass[12pt,amsmath,amssymb,floatfix]{iopart}

\usepackage{iopams,graphicx}
\begin{document}

\title{Prediction for new magnetoelectric fluorides}

\author{G. N\'{e}nert*, and T. T. M. Palstra}

\address{Solid State Chemistry Laboratory, Zernike Institute for Advanced Materials,\\ University of
Groningen, Nijenborg 4, 9747 AG Groningen, The Netherlands}

\begin{abstract}\\
We use symmetry considerations in order to predict new
magnetoelectric fluorides. In addition to these magnetoelectric
properties, we discuss among these fluorides the ones susceptible
to present multiferroic properties. We emphasize that several
materials present ferromagnetic properties. This ferromagnetism
should enhance the interplay between magnetic and dielectric
properties in these materials.
\end{abstract}

\maketitle

\section{Introduction}

In recent years, there has been a renewed interest in the
coexistence and interplay of magnetism and electrical polarization
\cite{Fiebig,Eerenstein,Maxim}. This interest has been
concentrated on multiferroics and magnetoelectric materials. In
multiferroics, a spontaneous polarization coexists with a long
range magnetic order. In magnetoelectrics (we consider here only
the linear effect), the polarization is induced by a magnetic
field in a magnetically ordered phase \cite{ITC}. In the Landau
theory framework, multiferroics which are not magnetoelectric
present at least a coupling of the type P$^{2}M^{2}$ (P:
polarization, M: total magnetization) while linear
magnetoelectrics are characterized by terms like PM$^{2}$ or LMP
(L: antiferromagnetic order parameter) \cite{Toledano}. Terms like
P$^{2}M^{2}$ are of higher degree than PM$^{2}$ or LMP terms.
Consequently, we expect a stronger interplay between dielectric
and magnetic properties in linear magnetoelectrics than in simple
multiferroics (e.g. YMnO$_{3}$ \cite{Agung}). More complicated
coupling terms can also characterize the magnetoelectric effect
(e.g. magnetic gradient)\cite{Harris}. These kind of terms are
outside the purpose of the present contribution. In the search for
materials presenting a strong coupling of magnetism and
polarization, the most promising ones are multiferroics presenting
linear magnetoelectric properties. These materials are scarce.
Thus, it is of interest to look for new magnetoelectric materials
by itself.

Recent efforts have concentrated on two main ideas: magnetic
frustration and breaking of the inversion center due to an
antiferromagnetic ordering. These approaches have been generated
by the ideas of on one side Katsura \cite{Katsura} and Sergienko
\cite{Sergienko} and on the other side of Mostovoy
\cite{Mostovoy}. They described in the case of non collinear
magnets a possible mechanism for magnetoelectricity and
polarization induced by antiferromagnetic ordering, respectively.
The new mechanism proposed by Katsura \emph{et al.} does not
involve the Dzialoshinskii-Moriya (DM) interaction contrary to
typical magnetoelectric compound such as Cr$_{2}$O$_{3}$
\cite{Cr2O3}. Most of the recent research on multiferroics
concerns centrosymmetric oxides \cite{nature}. These materials
present a breaking of the symmetry giving rise to a spontaneous
polarization which may be or not reversible by application of a
magnetic field. The idea of using symmetry analysis to predict
magnetoelectric compounds is not new. The first reported
magnetoelectric compound Cr$_{2}$O$_{3}$ was predicted to be
magnetoelectric prior to any experimental evidence
\cite{Dzialoshinskii}. It is the same philosophy that we aim to
take here.

In this article, we present a symmetry analysis of selected
materials. All these materials should present magnetoelectricity
based on symmetry arguments. We made a literature survey
considering various magnetically ordered compounds for which
neutron data were available. We made a systematic symmetry
analysis of all the studied compounds (about 50 materials). We
present here only our investigation of selected fluorides. This
choice is motivated by two reasons. The first one is that there is
a need to look for other materials than oxides if we search for
new materials since magnetoelectric/multiferroic materials are
scarce. The second reason is that polarization cannot exist in
conducting materials. Thus, the high charge transfer in the
fluorides make them good candidates for experimental
investigations.

Several fluorides were reported to crystallize in a polar
structure. Consequently, in addition to magnetoelectric
properties, several fluorides are potentially ferroelectric. When
this is the case, we discuss this possibility in the light of
known ferroelectrics related to the material under investigation.
All the compounds discussed in this article have been the subject
of detailed crystallographic and magnetic studies by means of
neutron diffraction. We present below the results of our search
for new magnetoelectric fluorides.

\section{Study of $\alpha$-KCrF$_{4}$}

\begin{figure}[htb]
\centering
\includegraphics[width=8cm]{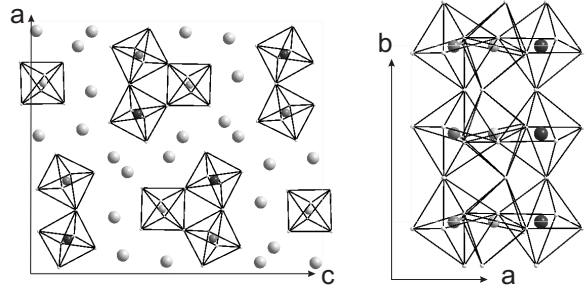}\\
\caption{Crystal structure of KCrF$_{4}$ projected along \emph{b}
left) and \emph{c} axis (right). We show the Cr$^{3+}$ sites in
their octahedral environment. The white atoms are the K$^{+}$
atoms. The different grey scales represent the three inequivalent
Cr$^{3+}$ sites.} \label{KCrF4-1}
\end{figure}

$\alpha$-KCrF$_{4}$ is the first in the selected fluorides we
present with possible magnetoelectric properties. The crystal
structure of $\alpha$-KCrF$_{4}$ is orthorhombic (space group
$Pnma$ (n$^{\circ}$62), \emph{a} = 15.76 \r{A}, \emph{b}= 7.43
\r{A}, \emph{c} = 18.38 \r{A}). It consists of infinite columns of
CrF$_{6}$ octahedra sharing edges along the \emph{b} axis (see
Fig. \ref{KCrF4-1}) \cite{kissel}.

\begin{figure}[htb]
\centering
\includegraphics[width=8cm]{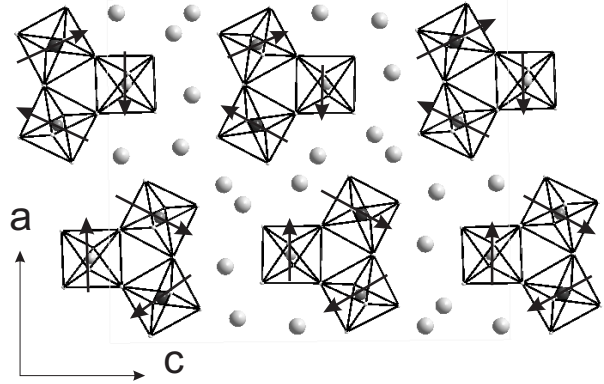}\\
\caption{Magnetic structure of KCrF$_{4}$ in the
(\emph{a},\emph{c}) plane. Arrows indicate the magnetic moments on
he chromium atoms with a quasi-120$^{\circ}$
configuration.}\label{KCrF4-2}
\end{figure}

This compound presents a high magnetic frustration among the
fluorides that we present here. It orders antiferromagnetically
only under T$_{N}$ = 4 K with a quasi 1D behavior. We present in
Fig. \ref{KCrF4-2} a representation of its magnetic structure as
determined from neutron scattering \cite{lacorre1}.

There are three inequivalent Cr$^{3+}$ ions per unit cell and
occupying the Wyckoff position 8d. Consequently, we have eight
different magnetic sites all carrying one spin S$_{j}$. We can
define the following eight magnetic vectors (one ferromagnetic and
seven antiferromagnetic ones):

\begin{eqnarray}
\label{1}
&\overrightarrow{M}=\overrightarrow{S_{1}}+\overrightarrow{S_{2}}+\overrightarrow{S_{3}}+\overrightarrow{S_{4}}+\overrightarrow{S_{5}}+\overrightarrow{S_{6}}+\overrightarrow{S_{7}}+\overrightarrow{S_{8}},\\
&\overrightarrow{L_{1}}=\overrightarrow{S_{1}}-\overrightarrow{S_{2}}+\overrightarrow{S_{3}}-\overrightarrow{S_{4}}+\overrightarrow{S_{5}}-\overrightarrow{S_{6}}+\overrightarrow{S_{7}}-\overrightarrow{S_{8}},\\
&\overrightarrow{L_{2}}=\overrightarrow{S_{1}}+\overrightarrow{S_{2}}-\overrightarrow{S_{3}}-\overrightarrow{S_{4}}+\overrightarrow{S_{5}}+\overrightarrow{S_{6}}-\overrightarrow{S_{7}}-\overrightarrow{S_{8}},\\
&\overrightarrow{L_{3}}=\overrightarrow{S_{1}}-\overrightarrow{S_{2}}-\overrightarrow{S_{3}}+\overrightarrow{S_{4}}+\overrightarrow{S_{5}}-\overrightarrow{S_{6}}-\overrightarrow{S_{7}}+\overrightarrow{S_{8}},\\
&\overrightarrow{L_{4}}=\overrightarrow{S_{1}}+\overrightarrow{S_{2}}+\overrightarrow{S_{3}}+\overrightarrow{S_{4}}-\overrightarrow{S_{5}}-\overrightarrow{S_{6}}-\overrightarrow{S_{7}}-\overrightarrow{S_{8}},\\
&\overrightarrow{L_{5}}=\overrightarrow{S_{1}}-\overrightarrow{S_{2}}+\overrightarrow{S_{3}}-\overrightarrow{S_{4}}-\overrightarrow{S_{5}}+\overrightarrow{S_{6}}-\overrightarrow{S_{7}}+\overrightarrow{S_{8}},\\
&\overrightarrow{L_{6}}=\overrightarrow{S_{1}}+\overrightarrow{S_{2}}-\overrightarrow{S_{3}}-\overrightarrow{S_{4}}-\overrightarrow{S_{5}}-\overrightarrow{S_{6}}+\overrightarrow{S_{7}}+\overrightarrow{S_{8}},\\
&\overrightarrow{L_{7}}=\overrightarrow{S_{1}}-\overrightarrow{S_{2}}-\overrightarrow{S_{3}}+\overrightarrow{S_{4}}-\overrightarrow{S_{5}}+\overrightarrow{S_{6}}+\overrightarrow{S_{7}}-\overrightarrow{S_{8}}
\end{eqnarray}

Lacorre and collaborators have investigated also the
transformation properties of the different components of the
magnetic vectors. We reproduce in Table \ref{KCrF4-IRs} the
results of their derivations \cite{lacorre1}.

\begin{table}[htb]
\centering
\begin{tabular}{|{c}|{c}|}
\hline
IR & Magnetic components\\
\hline
$\Gamma_{1}$ & L$_{1x}$, L$_{2y}$, L$_{3z}$  \\
\hline
$\Gamma_{2}$     & M$_{x}$, L$_{3y}$, L$_{2z}$\\
\hline
$\Gamma_{3}$     & L$_{2x}$, L$_{1y}$, M$_{z}$   \\
\hline
$\Gamma_{4}$     & L$_{3x}$, M$_{y}$, L$_{1z}$ \\
\hline
$\Gamma_{5}$     & L$_{5x}$, L$_{6y}$, L$_{7z}$  \\
\hline
$\Gamma_{6}$     & L$_{4x}$, L$_{7y}$, L$_{6z}$   \\
\hline
$\Gamma_{7}$     & L$_{6x}$, L$_{5y}$, L$_{4z}$  \\
\hline
$\Gamma_{8}$     & L$_{7x}$, L$_{4y}$, L$_{5z}$   \\
\hline
\end{tabular}
\\
\caption{Magnetic components classified by irreducible
representation.}\label{KCrF4-IRs}
\end{table}

As stated above, we need to look for the possible LMP terms
allowed by symmetry. These terms are the signature of the linear
magnetoelectric effect. For this, we need to know what are the
transformation properties of the polarization components. It is
sufficient to look at the transformation properties of the
different polarization components under the effect of the
generators of the space group. In Table \ref{Polarization-KCrF4},
we present the transformation properties of the polarization
components in the space group $Pnma$.

\begin{table}[htb]
\centering
\begin{tabular}{|{c}|{c}|{c}|{c}|}
\hline
& 2$_{1x}$ & 2$_{1z}$ & $\overline{1}$\\
\hline
P$_{x}$     &  1  &    -1  & -1  \\
\hline
P$_{y}$     & -1  &   -1  &  -1  \\
\hline
P$_{z}$     & -1  &    1  &  -1   \\
\hline
\end{tabular}
\\
\caption{Transformation properties of the polarization components
for the space group $Pnma1'$ associated to \textbf{k} =
0.}\label{Polarization-KCrF4}
\end{table}

According to the Tables \ref{KCrF4-IRs} and
\ref{Polarization-KCrF4}, we can determine the allowed LMP terms
which may be present and giving rise to an induced polarization
under magnetic field. We know that below T$_{N}$, the magnetic
structure is described by the irreducible representation
$\Gamma_{6}$. It is experimentally observed that
L$_{4x}$$>$L$_{6z}$ and L$_{7z}\simeq$0 \cite{lacorre1}. Taking
into account these experimental results, we find that the most
relevant magnetoelectric terms are L$_{4x}$P$_{y}$M$_{z}$ and
L$_{4x}$P$_{z}$M$_{y}$. Consequently, an induced polarization may
appear along P$_{y}$ (P$_{z}$) if one applies a magnetic field
along z (y). Since this compound is centrosymmetric, it cannot
present a multiferroic character.

\section{Study of KMnFeF$_{6}$}

\begin{figure}[htb]
\centering
\includegraphics[width=8cm]{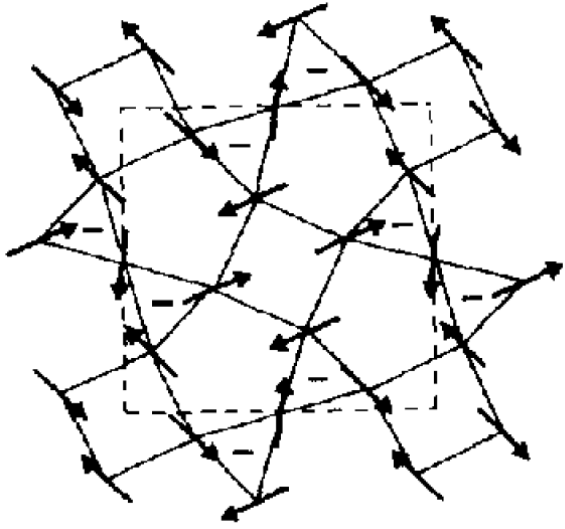}\\
\caption{Magnetic structure of KMnFeF$_{6}$ in the
(\emph{a},\emph{b}) plane. Arrows indicate the magnetic moments on
the iron atoms (mostly along the \emph{a} axis) from
\cite{lacorre2}.} \label{K2MnFeF6}
\end{figure}

The fluoride KMnFeF$_{6}$ presents a partial ordering of the Mn
and Fe atoms giving rise to an enlargement of the unit cell
compared to the usual tetragonal tungsten bronze type
\cite{lacorre2}. The family of tetragonal tungsten bronze and
related ones have been extensively investigated due to their
ferroelectric properties \cite{ferro}. This compound crystallizes
in the space group $Pba2$ (n$^{\circ}$32), where the Mn and Fe
ions order on the 8c Wyckoff position of the structure and occupy
statistically the 4b Wyckoff position. This compound is
magnetically frustrated due to the presence of triangular cycles
of antiferromagnetic interactions. All the Mn and Fe cations have
an octahedral environment of fluorine atoms. In the \emph{ab}
plane, Mn and Fe ions alternate along the \emph{c} axis. The
magnetic structure is presented in Fig. \ref{K2MnFeF6}
\cite{lacorre2}. Although the ferroelectric properties have not
been investigated to our knowledge, this compound is likely to
present a multiferroic character below T$_{C}$. Indeed since many
materials of this family are ferroelectric, it is likely that this
compound presents such property.

Although presenting magnetic frustration, the compound
KMnFeF$_{6}$ orders ferrimagnetically below T$_{C}$ = 148 K with a
ratio $\frac{\Theta}{T_{C}}$=3. The magnetic structure is
identical to the chemical unit cell and thus $\overrightarrow{k}$
= $\overrightarrow{0}$. The symmetry analysis by Bertaut's method
gives rise to the results presented in Table \ref{table1}
\cite{lacorre2,Bertaut}.

\begin{table}[htb]
\centering
\begin{tabular}{|{c}|{c}|{c}|{c}|{c}|}
\hline
Modes & x  & y & z & Magnetic space groups\\
\hline
$\Gamma_{1}$     & G$_{x}$  &   A$_{y}$  &   C$_{z}$ & $Pba2$ \\
\hline
$\Gamma_{2}$     & C$_{x}$  &   F$_{y}$  &   G$_{z}$ & $Pba'2'$ \\
\hline
$\Gamma_{3}$     & A$_{x}$  &   G$_{y}$  &   F$_{y}$ & $Pb'a'2$  \\
\hline
$\Gamma_{4}$     & F$_{x}$  &   C$_{y}$  &   A$_{y}$ & $Pb'a2'$  \\
\hline
\end{tabular}
\\
\caption{Irreducible representations for the space group $Pba21'$
associated to \textbf{k}=0.}\label{table1}
\end{table}

The neutron data show that the best model for the magnetic
structure is given by the $\Gamma_{4}$ mode. The corresponding
magnetic space group is thus $Pb'a2'$ which has the magnetic point
group m'm2'. According to Ref. 4, we have a linear magnetoelectric
effect which is allowed having the following allowed terms (after
transformation of the coordinates system):

\begin{displaymath}
\centering \label{2} \mathbf{[\alpha_{ij}]} = \left(
\begin{array}{ccc}
0 & 0 & 0 \\
0 & 0 & \alpha_{23} \\
0 & \alpha_{32} & 0 \\
\end{array} \right)
\end{displaymath}

We remind that KMnFeF$_{6}$ presents a polar structure and is
likely to be ferroelectric. Consequently, KMnFeF$_{6}$ is a
multiferroic material which presents a strong interplay between
magnetism and polarization below T$_{C}$=148K. Moreover, we notice
here that it would be one of the scarce ferrimagnetic compounds
presenting such properties. Under the application of a magnetic
field below T$_{C}$ along the \emph{c} axis (direction of
spontaneous polarization) should create a polarization along the
\emph{b} axis (term $\alpha_{23}$) and vice versa (term
$\alpha_{32}$). Thus it will be possible to switch the
polarization direction under the application of a magnetic field.
This is of high interest for technological applications. Another
remarkable feature is that this compound orders at 148K which is
much higher than the actual compounds \cite{nature}.

\section{Study of 2 members of the Ba$_{6}$M$_{n}$F$_{12+2n}$ family}

In the previous fluorides, the magnetic frustration appeared in
corner-sharing octahedra, which leads to a single type of
interaction. P. Lacorre and coworkers have been also investigating
compounds like Ba$_{2}$Ni$_{3}$F$_{10}$ (n = 9) and
Ba$_{2}$Ni$_{7}$F$_{18}$ (n = 21) which are members of the
Ba$_{6}$M$_{n}$F$_{12+2n}$ family \cite{lacorre3, lacorre4}. In
this family where M=Ni, there are not only corner-sharing
octahedra but also edge-sharing octahedra. Both types of
interaction exist in the Ba$_{2}$Ni$_{3}$F$_{10}$ and
Ba$_{2}$Ni$_{7}$F$_{18}$ compounds. These compounds have been
investigated by means of powder neutron diffraction at room and
low temperatures.

We start by looking at the Ba$_{2}$Ni$_{3}$F$_{10}$ material. This
compound crystallizes in the space group $C2/m$ (n$^{\circ}$12)
containing 3 different Ni$^{2+}$ per unit cell. 2 Ni ions occupy
the Wyckoff position 4i and the other one occupies the Wyckoff
position 4h. Below T$_{N}$ = 50 K, an antiferromagnetic ordering
starts to develop characterized by a magnetic wave-vector
$\overrightarrow{k}$=(0,0,1/2). All the (hkl) magnetic reflections
do not satisfy the C-centering of the chemical cell but a
primitive lattice. P. Lacorre and collaborators have shown that
the magnetic space group is $P2/m'$ where the magnetic moments lie
in the \emph{ac} plane. Consequently, the magnetic point group of
this compound below its T$_{N}$ is 2/m'. According to Ref. 4, a
linear magnetoelectric effect is allowed having the following
expression:

\begin{displaymath}
\centering \label{2} \mathbf{[\alpha_{ij}]} = \left(
\begin{array}{ccc}
\alpha_{11}  &       0      & \alpha_{13} \\
         0   &  \alpha_{22} &          0  \\
\alpha_{31}  &     0        & \alpha_{33}  \\
\end{array} \right)
\end{displaymath}

Consequently, induced polarization can be observed along the three
crystallographic directions under the application of an applied
magnetic field. This material is not multiferroic since its
structure is centrosymmetric. Moreover the structure remains
centrosymmetric in the magnetic ordered phase. Consequently no
spontaneous polarization can develop below and above T$_{N}$.

The other member of the family of interest is for n=21.
Ba$_{2}$Ni$_{7}$F$_{18}$ crystallizes in the polar space group
$P1$ (n$^{\circ}$1) containing four inequivalent sets of Ni$^{2+}$
ions. Each Ni$^{2+}$ ion occupies the Wyckoff position 1a in the
general position. From all the fluorides that we treat here, it is
the second which orders ferrimagnetically under T$_{C}$ = 36 K.
Due to the low symmetry of the crystal, we have to deal here with
magnetic components along the three crystallographic directions.
While all the already studied fluorides present magnetic
frustrations, it is not the case in this compound. We mean there
is no competition between next nearest neighbors. Below T$_{C}$,
all the new magnetic reflections can be indexed in the same cell
as the chemical one. Consequently, the star of the magnetic
wave-vector has only one arm. The irreducible representations
associated to the space group $P1$ with
$\overrightarrow{k}$=$\overrightarrow{0}$ are given in Table
\ref{P11}.

\begin{table}[htb]
\centering
\begin{tabular}{|{c}|{c}|}
\hline
& h$_{1}$  \\
\hline
 $\Gamma_{1}$     & 1 \\
\hline
\end{tabular}
\\
\caption{Irreducible representation for the space group $P11'$
associated to \textbf{k}=(0, 0, 0).}\label{P11}
\end{table}

According to the Table \ref{P11}, there is only one possibility
for the magnetic space group which is $P1$. Referring to the Ref.
4, a linear magnetoelectric effect is allowed with non-zero
components:

\begin{displaymath}
\centering \label{2} \mathbf{[\alpha_{ij}]} = \left(
\begin{array}{ccc}
\alpha_{11}  &  \alpha_{12} &  \alpha_{13} \\
\alpha_{21}  &  \alpha_{22} &  \alpha_{23} \\
\alpha_{31}  &  \alpha_{32} &  \alpha_{33}  \\
\end{array} \right)
\end{displaymath}

Consequently, Ba$_{2}$Ni$_{7}$F$_{18}$ is a potential multiferroic
material (polar structure and ferrimagnetic below T$_{C}$=36K).
Moreover, irrespective of the direction of an applied magnetic
field, the polarization parallel to the magnetic field will
increase due to the magnetoelectric effect below T$_{C
}$.

\section{Study of CsCoF$_{4}$}

\begin{figure}[htb]
\centering
\includegraphics[width=8cm]{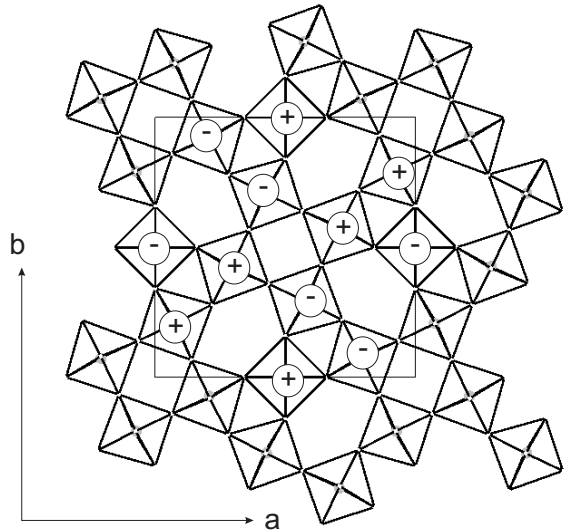}\\
\caption{Magnetic structure of CsCoF$_{4}$  in the
(\emph{a},\emph{b}) plane. Plus and Minus signs indicate the
magnetic moments along the \emph{c} axis (up or down).}
\label{CsCoF4}
\end{figure}

CsCoF$_{4}$ is the last compound among the fluorides that we
investigate in the light of a possible magnetoelectric effect.
This compound crystallizes in the non-polar space group
$I\overline{4}c2$ (n$^{\circ}$120) with two different Co$^{3+}$
Wyckoff positions in the unit cell: 4d and 16i. The
antiferromagnetic order occurring below T$_{N}$ = 54 K is
characterized by a magnetic wave-vector $\overrightarrow{k}$ =
$\overrightarrow{0}$ \cite{lacorre5}. This structure is also
magnetically frustrated due the presence of ferromagnetic
interactions within an antiferromagnetic plane as described in
Fig. \ref{CsCoF4}.

Based on geometrical considerations and comparison with magnetic
structure of compounds of the same family (namely LiCoF$_{4}$),
the authors proposed some constraints on the orientation of the
magnetic moments. From these considerations, they found that the
magnetic space group of CsCoF$_{4}$ is $I\overline{4}'$. The
corresponding magnetic point group is $\overline{4}'$. If one
compares this magnetic point group with the ones listed in Ref. 4,
we observe that a linear magnetoelectric effect is possible along
several directions:

\begin{displaymath}
\centering \label{2} \mathbf{[\alpha_{ij}]} = \left(
\begin{array}{ccc}
 \alpha_{11}  & \alpha_{12} & 0 \\
-\alpha_{12}  & \alpha_{11} & 0 \\
           0  &     0       & \alpha_{33}  \\
\end{array} \right)
\end{displaymath}

\section{Discussion}

In the previous sections we have investigated the magnetic
symmetry of various fluorides. We shall discuss here the common
mechanism which may give rise to the magnetoelectric effect in the
studied fluorides and compare it to other known magnetoelectric
fluorides such as BaMnF$_{4}$ \cite{BaMnF4}. But prior to this, we
should stress that there is an upper bound for the magnetoelectric
effect \cite{Brown} which is defined as:

\begin{eqnarray}
\centering \label{2}
\alpha_{ij}\leq\varepsilon_{0}\mu_{0}\varepsilon_{ii}\mu_{jj}
\end{eqnarray}

$\varepsilon_{ii}$ and $\varepsilon_{0}$ are respectively the
permittivity of free space and the relative permittivity of the
considered material. While $\mu_{jj}$ and $\mu_{0}$ are the
relative permeability and the permeability of free space,
respectively. As a consequence of Eq. (\ref{2}), the
magnetoelectric effect will remain small compared to unity except
possibly in ferroelectric and ferromagnetic materials. Thus
multiferroic magnetoelectric materials with ferromagnetic order
are the most interesting. Among the various compounds that we have
investigated, KMnFeF$_{6}$ and Ba$_{2}$Ni$_{7}$F$_{18}$ are likely
to be good representatives of such materials.

Various mechanisms may contribute to the magnetoelectric effect.
In the old literature, we can count about four different
mechanisms which may participate in the magnetoelectric effect
\cite{alcantara}. We can consider the molecular field theory
expression of a magnetic field for the magnetoelectric
susceptibility.

\begin{eqnarray}
\centering \label{3} H=H_{0}+V\
\end{eqnarray}

where the Hamiltonian H$_{0}$ describes the spin system in the
presence of a magnetic field and the perturbation V is linear in
the electric field.

\begin{eqnarray}
\centering \label{4}
H_{0}=\frac{1}{2}\sum_{ij}J_{ij}\mathbf{S}_{i}.\mathbf{S}_{j}-D\sum_{j}\left(S_{i}^{z}\right)^{2}-\mu
\mathbf{H}.\sum_{i}\mathbf{S}_{i}
\end{eqnarray}

V represent the changes of the various tensors due to the presence
of an electric field: single-ion anisotropy, g-factor, symmetric
exchange and Dzyaloshinskii-Moriya interactions. For the detailed
expression of V, we refer the reader to the literature (see Ref.
25). It has been shown that in the presence of an electric field
the changes of the g factor are predominant at low temperature in
GdAlO$_{3}$ compared to the changes for the other tensors but not
anymore above 1.2 K (T$_{N}$=3.78 K) \cite{alcantara}. It is thus
difficult to determine which parameter has the most important
contribution since it depends not only on the compound but also on
the temperature. The change of the anisotropy energy, exchange and
g value due to the electric field have been proposed for the
origin of the magnetoelectric effect in Cr$_{2}$O$_{3}$
\cite{Cr2O3}. These various mechanisms are susceptible to play a
role in the magnetoelectric effect in the fluorides that we
present in the previous sections.

The main difference between the various fluorides that we present
is the presence of inversion symmetry. The Dzyaloshinskii-Moriya
(DM) interaction (antisymmetric exchange) is allowed only when the
inversion symmetry is broken at the ligand ion mediating the
exchange \cite{DM}. Therefore, when the crystal structure has
inversion symmetry, the external electric field \textbf{E} induces
the DM interaction. Thus the DM vector can be defined as
\textbf{D}$_{ij}\propto$\textbf{E}$\times$\textbf{e}$_{ij}$ with
\textbf{e}$_{ij}$ being the unit vector connecting the two sites i
and j. This is the mechanism which has been proposed to explain
the magnetoelectric effect in ZnCr$_{2}$Se$_{4}$ \cite{ZnCr2Se4}.
One may expect that the contribution of the DM mechanism to the
magnetoelectric effect is higher for compounds which allows a
spontaneous DM interaction (i.e. not induced by the electric field
\textbf{E}). In this perspective, we expect that the almost
collinear magnetic structure of CsCoF$_{4}$ will give rise to a
negligible DM contribution to the predicted magnetoelectric
effect. We note that DM interactions do not result systematically
in a magnetoelectric contribution as in the case of
$\alpha$-Fe$_{2}$O$_{3}$ \cite{DM} or CoF$_{2}$ \cite{CoF2}, which
are piezomagnetic materials.

The fluorides that we present in this contribution present various
symmetry properties. We can classify them in two types of crystal
structures: polar and non-polar. In the first category, we count
KMnFeF$_{6}$ and Ba$_{2}$Ni$_{7}$F$_{18}$. In the other category,
we have KCrF$_{4}$, Ba$_{2}$Ni$_{3}$F$_{10}$ and CsCoF$_{4}$. As a
consequence of their polar structure, KMnFeF$_{6}$ and
Ba$_{2}$Ni$_{7}$F$_{18}$ are potentially multiferroic and thus
ferroelectric at room temperature. If their ferroelectric
properties can be confirmed experimentally, the mechanism for this
ferroelectricity remains to be investigated. Several multiferroic
fluorides have been investigated theoretically and experimentally
\cite{BaMnF4, fluorides}. It has been shown that the typical
charge transfer towards empty d-orbitals responsible for
ferroelectricity such as in BaTiO$_{3}$ is not active in
BaMF$_{4}$ (M = Mn, Fe, Co, and Ni). The ferroelectric instability
in the multiferroic barium fluorides arises solely due to size
effects \cite{fluorides}. In the light of the occupied d-orbitals
in KMnFeF$_{6}$ and Ba$_{2}$Ni$_{7}$F$_{18}$, it could be that the
cooperative displacements of K$^{+}$ and Ba$^{2+}$ respectively
would be responsible for the ferroelectric instability. Obviously,
this hypothesis remains to be confirmed experimentally.

\section{Conclusion}

In conclusion, we have shown from symmetry analysis that several
fluorides are likely to be magnetoelectric. Several of them may
present a multiferroic character coupled to an induced
polarization under the application of a magnetic field. Most of
them present magnetic frustration. We present here possible
magnetoelectrics which are among the scarce ferrimagnetic systems.
This ferromagnetism may enhance the interplay between polarization
and magnetism for the case of multiferroic materials. The
mechanism for potential ferroelectricity and magnetoelectric
effect remain to be investigated. We expect that this work will
stimulate experimental investigations of the dielectric properties
of the above reported fluorides.

\section*{ACKNOWLEDGEMENTS}
The work was supported by the Dutch National Science Foundation
NWO by the breedtestrategieprogramma of the Materials Science
Center, MSC$^{+}$.

\end{document}